\documentclass[twocolumn,prb,showpacs,preprintnumbers,superscriptaddress,amsmath,amssymb,citeautoscript]{revtex4-1}
\usepackage{graphicx}
\usepackage{dcolumn}
\usepackage{color}
\usepackage{bm}
\usepackage[hidelinks]{hyperref}
\hypersetup{
    colorlinks,
    citecolor=blue,
    filecolor=blue,
    linkcolor=blue,
    urlcolor=blue
}

\newcommand{\ud}{\,\mathrm{d}}
\newcommand{\im}{\textrm{i}}

\DeclareMathOperator{\C}{\mathcal{C}}
\DeclareMathOperator{\U}{\mathcal{U}}
\DeclareMathOperator{\M}{\mathcal{M}}

\DeclareMathOperator{\V}{\mathcal{V}}
\DeclareMathOperator{\T}{\mathcal{T}}
\DeclareMathOperator{\Order}{O}

\begin{document}

\title{Majoranas with and without a `character': hybridization, braiding\\ and Majorana number}

\author{N. Sedlmayr}
\email{nsedlmayr@hotmail.com}
\author{M. Guigou}
\affiliation{Institut de Physique Th\'eorique, CEA/Saclay,
Orme des Merisiers, 91190 Gif-sur-Yvette Cedex, France}
\author{P. Simon}
\affiliation{Laboratoire de Physique des Solides, UMR 8502, B\^at. 510, 91405 Orsay Cedex, France}
\author{C. Bena}
\affiliation{Institut de Physique Th\'eorique, CEA/Saclay,
Orme des Merisiers, 91190 Gif-sur-Yvette Cedex, France}
\affiliation{Laboratoire de Physique des Solides, UMR 8502, B\^at. 510, 91405 Orsay Cedex, France}

\date{\today}

\begin{abstract}
In this paper we demonstrate under what conditions a pseudo-spin degree of freedom or character can be ascribed to Majorana bound states (MBS). These exotic states can be created at the boundaries of non-interacting systems, corresponding to D, DIII and BDI in the usual classification scheme, and we focus on one dimension. 
We have found that such a character is  directly related to the class of the topological superconductor and its description by a $\mathbb{Z}$, rather than a $\mathbb{Z}_2$, invariant which corresponds to the BDI class.
We have also found that the DIII case with mirror symmetry, which supports multiple MBS, is in fact equivalent to the BDI class with an additional time-reversal symmetry. In all cases where a character can be given to the Majorana states we show how to construct the appropriate local operator
explicitly with various examples.  We also examine the consequences of the Majorana character by considering possible hybridization of MBS brought into proximity and find that two MBS with the same character do not hybridize.
Finally, we show that having this character or not has no consequence on the braiding properties of MBS.
\end{abstract}


\maketitle

\section{Introduction}

In his seminal 2001 paper, Kitaev introduced a simple toy model of a 1d topological p-wave superconductor of length $L$ \cite{Kitaev2001}.
Local Majorana operators were introduced by decomposing each local  annihilation and creation electronic operators $c_j$ and $c^\dagger_j$ at a given site $j$ into a pair of local Majorana operators
 defined by $\gamma_{a,j}=(c_j+c^\dagger_j)/2$, $\gamma_{b,j}=(c_j-c^\dagger_j)/2i$. The topological phase is characterized by two localized Majorana fermions at both extremities of the chain. When the pairing and hopping amplitudes are equal, the Majorana operators $\gamma_{a,1}$ and $\gamma_{b,L}$ decouple from the Hamiltonian. These are the zero-energy modes characterized by Majorana operators $\gamma_{a,1}$ and $\gamma_{b,L}$. Although in this decomposition, the  indices $a,b$ may appear somehow arbitrary (note also that $a$ and $b$ can be exchanged), they are related to the trivial feature that a Majorana operator is always built from two different fermionic operators in any realistic condensed matter Hamiltonian.  One may wonder whether - and under which conditions - we can associate to this label a kind of pseudo-spin degree of freedom or `character'. In the Kitaev chain, if a Majorana fermion of type `a' is localized at one end of the chain, then necessarily a Majorana fermion of type `b' is localized at the other extremity. In this paper, we will be focusing on non-interacting 1d systems.

In order to attribute some physical meaning to this `character', we are looking for a local operator $\V$ which (i) anti-commutes with the Hamiltonian $H$, $\{H,\V\}=0$, (ii) obeys $\V^2=1$, and (iii) is unitary and thus {\it distinct} from the particle-hole symmetry operator. Such an operator is an example of a chiral symmetry \cite{Ryu2010}. The first two properties ensure that the Majorana bound states which can appear can be characterized as eigenstates of $\V$ with two possible eigenvalues defining the aforementioned `character' of the Majorana fermion. 
Moreover, it follows that if the Hamiltonian has reflection symmetry, then these Majorana bound states will be localized at the two boundaries.  Any continuous change to the Hamiltonian which does not close the gap and leaves the symmetry intact will preserve this. Thus we always have localized Majoranas with a definite character, provided the above conditions are satisfied.
 
Since we focus on 1d particle-hole symmetric non-interacting systems, they belong to one of the following three classes: BDI, D or DIII depending on the presence or absence of different anti-unitary symmetries \cite{Schnyder2008}. These symmetries are referred to as time-reversal symmetries and at the single particle level satisfy $\T_\pm^2=\pm1$ for BDI/DIII and are absent for D. In systems belonging to the topological class BDI, the symmetry associated to the operator $\V$ is the composite of the time-reversal and particle-hole symmetries and is also known as the chiral symmetry, the existence of which is already well-known \cite{Ryu2010}. We prove the equivalence between the chiral symmetry and our character operator for this class. The character however can also be viewed as a local property of the MBS, a property we will extensively exploit to analyze various examples.
Each `type' of Majorana is localized on a different boundary of the system. In the class D, one no longer has the symmetry which allows this classification. In 1d the existence of this character can be directly related to the change in topological invariant from $\mathbb{Z}$ for BDI to $\mathbb{Z}_2$ for class D as the possibility of having many Majorana states at the end of a wire \cite{Tewari2012} requires them to have a well-defined character. This is because, as we will demonstrate, Majorana states without such a character generically destroy each other on contact, recombining to form finite energy Dirac particles.

DIII topological superconductors are time-reversal invariant and have a $\mathbb{Z}_2$ invariant \cite{Zhang2013a,Zhang2013,Wong2012,Liu2014}. Thus generically they behave like class D but with Kramer's pairs of Majorana states. However in addition to time-reversal symmetry they can also still have a so called mirror symmetry $\M$ where $\M^2=-1$ \cite{Zhang2013a}. We show that when $[H,\M]=0$ one also has $\{H,\V\}=0$ and a character can be defined.
In these cases the same conclusions for the BDI class can be applied to the Kramer's pairs of Majorana bound states and one can have many Majorana states. When this mirror symmetry is broken, then $\{H,\V\}\neq0$ and the DIII class behaves like class D, which are both described by a $\mathbb{Z}_2$ invariant. The possible existence of the many Majorana states in the presence of the mirror symmetry indicates that it should have a $\mathbb{Z}$ invariant \cite{Zhang2013a}, and we show that this case is in fact equivalent to the BDI class with an additional time-reversal symmetry which does not change the symmetry class. Note that we are not arguing simply that the low energy sector is BDI, as can arise in some D systems \cite{Tewari2012}, but rather that in this case the full symmetry of the Hamiltonian is BDI, and not DIII, {\it without} further approximation. It has already been noted that multiple particle-hole symmetries or time reversal symmetries of the same type do not alter the classification scheme as the multiple of, for example, two time-reversal operators is a unitary operator \cite{Ryu2010}. However this does not in itself decide which class a system with two different types of time reversal symmetries belongs to, and consequently what its topological invariant is. The simultaneous presence of two {\it different} time-reversal symmetries is still ambiguous in the Schnyder et al.~classification scheme \cite{Schnyder2008,Ryu2010},  and we argue here that this class should be properly thought of as a subset of BDI.  We prove that although the DIII class also possesses a chiral symmetry this can not be related to a character.

The interest in Majorana bound states is partially fuelled by their possible application to quantum computing \cite{Kitaev2001,Nayak2008}. To this end the non-abelian braiding properties of the Majoranas are key \cite{Ivanov2001,Alicea2011,Wu2014}.  We consider the implications for the braiding properties for Majorana states which have no well defined character, where the usual mapping to a Kitaev chain is modified.
Majorana states with definite character can be shown to either hybridize to finite energy states, or to be protected and remain at zero energy. Which of these occurs depends on whether the Majoranas have the same character, when they are therefore protected, or a different character when they will hybridize. Majorana states which exist without character are much less robust and their mutual interaction will, unless one can carefully tune the appropriate overlaps to zero, destroy the zero energy Majorana states. The resulting finite energy states remain localized at the edges, but do not retain the Majorana anti-commutation relations which occur uniquely at zero energy. We go on to demonstrate that the character or absence of it has no known effects for the braiding properties of the Majorana states. Indeed, as one wishes for the states which are to be braided to hybridize between themselves, thus avoiding problems caused by degeneracies at zero energy, the case where the full system allows a well defined character needs to be avoided.

The paper is organized as follows. In Sec.~\ref{sec_charac}, we examine under which symmetry conditions we can build the Majorana character.
In Sec.~\ref{sec_hybrid} we demonstrate the consequences of the Majorana character by considering possible hybridization and destruction of Majorana bound states brought into proximity. In Sec.~\ref{braiding} we consider the implications of Majorana character or its absence on their braiding properties.
We then summarize our results and offer some perspectives in Sec.~\ref{conclusions}.

\section{Characterizing Majorana states }\label{sec_charac}

Before characterizing the Majorana states, let us introduce our notations and conventions.

\subsection{Particle-hole symmetry operator and notations}\label{sec_background}
We will use the Nambu basis throughout the paper:  $\Psi^\dagger_j=\{c^\dagger_{j\uparrow},c^\dagger_{j\downarrow},c_{j\downarrow},-c_{j\uparrow}\}$, where $c_{j\sigma}^{(\dagger)}$ annihilates (creates) a particle of spin $\sigma$ at a site $j$. The corresponding wavefunction is $\psi^T_j=\{u_{j\uparrow},u_{j\downarrow},v_{j\downarrow},v_{j\uparrow}\}$. We will also use $\vec{\bm\tau}$ to denote the Pauli matrices in the particle-hole subspace and $\vec{\bm\sigma}$ as the Pauli matrices in the spin subspace. We confine ourselves to discussing only generic Bogoliubov-de-Gennes Hamiltonians, $H$, which are antisymmetric under the particle-hole transformation
$\C={\rm e}^{\im\zeta}{\bm \sigma}^y{\bm \tau}^y{\hat K}$,  
i.e.~$\{H,\C\}=0$ and $\C^2=1$. Here $\hat K$ is the complex-conjugation operator, and $\zeta$ is an arbitrary phase which, without loss of generality, we will set to zero for convenience. Therefore the Hamiltonians have spectra $(\pm\epsilon_1,\pm\epsilon_2,\ldots,\pm\epsilon_{\frac{d}{2}})$, where $d$ is the Hilbert space dimension.

Now consider two eigenstates of the Hamiltonian $\xi_{\pm 1}\equiv\xi(\pm\epsilon_1)$ such that $(H\mp\epsilon_1)\xi_{\pm 1}=0$. Let
\begin{equation}\label{state1}
\xi_1=\sum_{j\sigma}[u_{j\sigma}c_{j\sigma}-\sigma v_{j\sigma}c^\dagger_{j\sigma}]\,,
\end{equation}
then by applying the particle-hole transformation
\begin{equation}\label{state2}
\xi_{-1}=\sum_{j\sigma}[-\sigma v^*_{j\sigma}c_{j\sigma}+u^*_{j\sigma}c^\dagger_{j\sigma}]\,.
\end{equation}
Here $\sigma=\uparrow,\downarrow$ should be taken as $\sigma=\pm 1$ whenever a numerical value is required. If the system is in a topologically non-trivial phase then Majorana states will be present at the boundaries \cite{Ryu2002,Sato2009a,Sato2009b,Lutchyn2010,Oreg2010}. In the thermodynamic limit, system size $L\to\infty$, the energy of these states goes to zero exponentially, $\lim_{L\to\infty}\epsilon_1\sim{\rm e}^{-L/\ell}$, where $\ell$ is a constant. It is then always possible to construct two Majorana states out of these degenerate solutions. These Majorana states can be localized at each boundary, though naturally other bases are possible. As a Majorana is its own anti-particle then a Majorana must be an eigenstate of $\C$ with an eigenvalue of magnitude $1$, $\C|\gamma\rangle={\rm e}^{\im\tilde\zeta}|\gamma\rangle$. Once again we can always tune the phase to be $\tilde\zeta=0$ by changing the irrelevant overall phase of the eigenstates. Therefore if we calculate the expectation value, with $\gamma^T\equiv\{u_{j\uparrow},u_{j\downarrow},v_{j\downarrow},v_{j\uparrow}\}=\{u_{j\uparrow},u_{j\downarrow},u_{j\downarrow}^*,-u_{j\uparrow}^*\}$, then we find
\begin{equation}
|<\gamma|\C|\gamma>|=|2\sum_{j,\sigma}\sigma u_{j\sigma}v_{j\sigma}|=|2\sum_{j,\sigma}\sigma|u_{j\sigma}|^2|=1\,.
\end{equation}
This is equivalent to imposing the appropriate anti-commutation relations on the Majorana states.
All non-zero energy eigenstates must satisfy
\begin{equation}
<\xi_n|\C|\xi_n>=<\xi_n|\xi_{-n}>=0\,.
\end{equation}

\subsection{General symmetry considerations}

The systems we consider generically possess the property that $\{H,\C\}=0$ with $\C^2=1$. 
We assume that  we also have some $\V$ which satisfies $\{H,\V\}=0$ for a given Hamiltonian $H$ with $\V^2=1$. Although the condition $|\V|^2=1$ would be sufficient, as $\V$ and $\C$ must commute the stronger condition $\V^2=1$ is required. Additionally the operator $\V$ should commute with all other symmetry operators of the Hamiltonian. As the operators $\C$ and $\V$ are assumed to be distinct and we have $\C\lambda=\lambda^*\C$ where $\lambda$ is a complex number and $\lambda^*$ the complex conjugate, $\V$ must satisfy $\V\lambda=\lambda\V$.
Now let us suppose we have two zero energy states $\xi_{\pm 1}$, where as before for a finite system zero energy really means exponentially small in the system size.  These states are related by $\C\xi_{\pm 1}=\xi_{\mp 1}$ and $\V\xi_{\pm 1}=\xi_{\mp 1}$, however $\C\im\xi_{\pm 1}=-\im\xi_{\mp 1}$ and $\V\im\xi_{\pm 1}=\im\xi_{\mp 1}$. Constructing the orthogonal Majorana states $\gamma_{1}=(\xi_{1}+\xi_{- 1})/\sqrt{2}$ and $\gamma_{2}=\im(\xi_{1}-\xi_{-1})/\sqrt{2}$ we immediately have that $\C\gamma_{1,2}=\gamma_{1,2}$ but that $\V\gamma_1=\gamma_1$ and $\V\gamma_2=-\gamma_2$. The two eigenvalues $\pm 1$ define the Majorana character.

There are three symmetry classes which we must consider. Firstly we have class BDI and class DIII which have a time-reversal invariance operator, $\T_\pm$, satisfying $[H,\T_\pm]=0$ and $\T_\pm^2=\pm1$ respectively. Then we have class D which has no time-reversal symmetry.

\subsubsection{Class BDI}
For the BDI class, the anti-commutation relation $\{H,\V\}=0$ and $\V^2=1$ are ensured for $\V$ defined by $\V=\C \T_+$. $\V$ is sometimes called the chiral or sub-lattice symmetry. This symmetry allows one to assign a chiral character to the Majoranas as we shall see, and this in turn allows many Majorana states to exist at the end of a wire without recombining. This is in agreement with such a system having a $\mathbb{Z}$ invariant. In the simplest cases $\T_+=K$ the complex conjugation operator and many other BDI cases can be transformed to this one by a global transformation, see App.~\ref{app_planar}.

\subsubsection{Class D}
For class D, we find that it is not possible to assign a character to the Majorana states and only a single Majorana can exists at one end of a topological wire. Such a system has a $\mathbb{Z}_2$ invariant. 

To prove this, let us assume that we can define an operator  $\V$ that satisfies our requirements, namely: $\{H,\V\}=[\C,\V]=0$ and $\V^2=1$. 
Then we can define an operator $\T=\C\V$. We see that such operator $\T$ must obey $[H,\T]=0$ and $\T^2=1$. Therefore this means we are in the BDI class and not in the D class  as originally assumed. Therefore this proves that class D has no character. Physically this makes sense if we regard the Majorana character as allowing the coexistence of other Majorana bound states. As this is impossible in class D in 1d due to the $\mathbb{Z}_2$ topological invariant, 
there cannot be any character in class D.

\subsubsection{Class DIII}
For class DIII we shall consider two cases. Firstly if we do not have mirror symmetry then the system is in the same situation as for the D class and we still have no character. The same proof as for the D case can be given. We note that an operator defined by $\bar\V=\C\T_-$ can not be used as $\bar\V^2=-1$. 

The second DIII case corresponds to the Hamiltonian having an extra-symmetry, a so-called mirror symmetry introduced in Ref.~\onlinecite{Zhang2013a}.
Mirror symmetry is defined by a unitary operator $\M$ which commutes with the Hamiltonian, $\C$ and $\T_-$ and also has $\M^2=-1$ \cite{Zhang2013a}. This is in fact a Hamiltonian with the symmetry $[H,\T_+]=0$ in addition to the physical time-reversal invariance of the DIII class $[H,\T_-]=0$. 
That mirror symmetry and simultaneous $\T_\pm$ symmetries are equivalent one can easily see by the relation $\T_+=\T_-\M$. It follows directly from the properties of $\T_-$ and $\M$ that $[H,\T_+]=0$ and $\T_+^2=1$, which are of course the definition of $\T_+$. Similarly, if one starts with $\T_+$ and $\T_-$, it then naturally follows that one can construct an operator $\M$ where $[H,\M]=0$ and $\M^2=-1$. The operator $\V$ remains the same as for the BDI class and all properties of these states transfer to the Kramer's pairs. As before $\V=\C\T_+$. 

This demonstration shows that a DIII system with a mirror symmetry should be thought of as a member of the BDI class instead.
As multiple Majorana states are possible, it must be characterized by a $\mathbb{Z}$ and not a $\mathbb{Z}_2$ invariant, as was  recognised early in Ref.~\onlinecite{Zhang2013a}.

\section{Consequences of the chiral Majorana character on hybridization}\label{sec_hybrid}

\begin{figure}\begin{center}
\includegraphics[width=0.95\columnwidth]{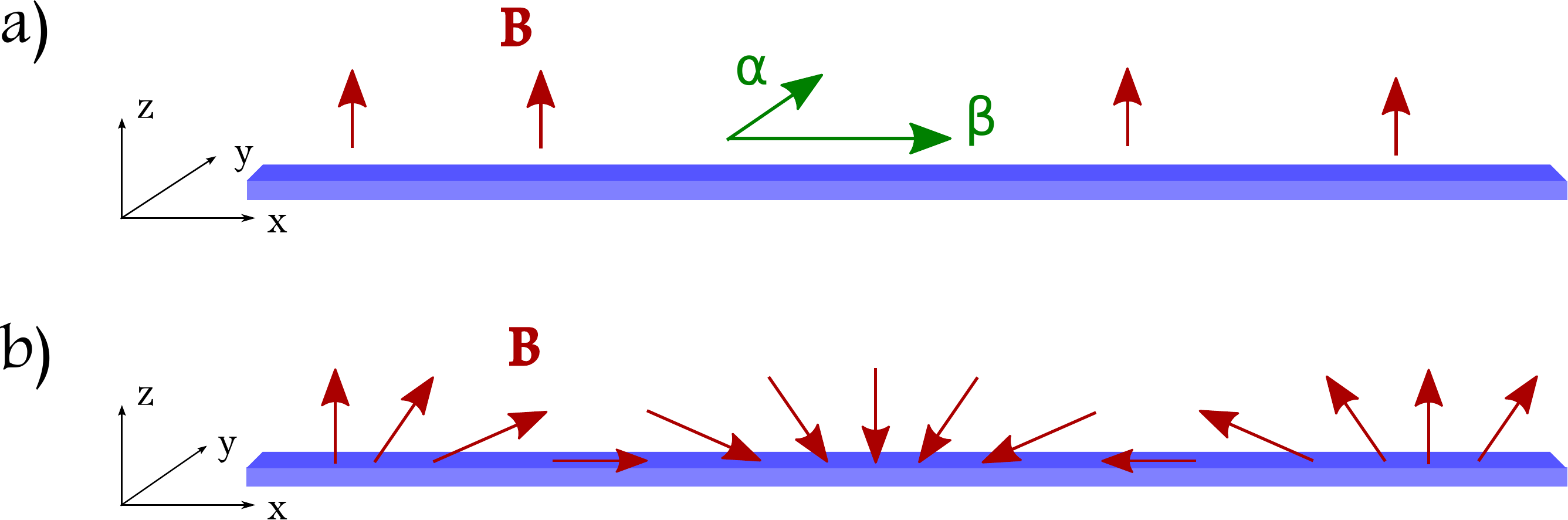}
\end{center}
\caption{(Color online) Schematics of a superconducting wire a) in the presence of a perpendicular homogeneous magnetic field $B$ and spin-orbit coupling $\alpha$ and $\beta$, b) in the presence of a rotating magnetic field $B$.}
\label{system}
\end{figure}

To investigate the robustness of the Majorana states at the ends of a wire we will consider a ladder made by two coupled wires, $H_\nu$ with $\nu=1,2$, described by the Hamiltonian
\begin{equation}
{\bm H}=\left(\begin{array}{cc}
H_1 & H_c\\
H_c^\dagger & H_2
\end{array}\right)\,.
\end{equation}
$H_c$ is the coupling between the wires. 
As a first step we diagonalize $H_{1,2}$ using
\begin{equation}
{\bm N}=\left(\begin{array}{cc}
N_1&0\\
0&N_2
\end{array}\right)
\end{equation}
where $N_{1,2}^\dagger H_{1,2} N_{1,2}$ are diagonal. For $H_\nu$ we introduce a general one dimensional tight-binding Hamiltonian on $L$ sites
\begin{equation}\label{hamiltonian}
H_\nu=H_0+H_{\rm B}+H_{\rm so}+H_{\rm s}+H_{\rm d}\,,
\end{equation}
There are already several possible experimental realizations of Majorana fermions, though definitive proof is still lacking \cite{Mourik2012,Deng2012,Das2012,Lee2014,Nadj-Perge2014}. The models introduced here are sufficient to capture the essential physics of these various systems. In Fig.~\ref{system} schematics of the wire systems are presented.

Suppressing for the moment the wire index $\mu$, the hopping and chemical potential are
\begin{equation}
H_0=\sum_j\left\{\Psi^\dagger_j\left[(\mu-t){\bm\tau}^z\right]\Psi_j
-\frac{t}{2}\Psi^\dagger_j{\bm \tau}^z\Psi_{j+1}+{\rm H.c.}\right\}\,.
\end{equation}
We consider two different forms of superconductivity induced by proximity to a substrate. Firstly we have s-wave superconductivity, for the BDI and D topological superconductors,
\begin{equation}
H_{\rm s}=-\sum_j\Psi^\dagger_j{\bm\Delta}^{\rm s}_j\Psi_j\,.
\end{equation}
For DIII we consider d-wave superconducting pairing induced by proximity with a $d_{x^2-y^2}$ superconducting substrate \cite{Wong2012}
\begin{equation}
H_{\rm d}=-\frac{1}{2}\sum_j\left[\Psi^\dagger_j{\bm\Delta}^{\rm d}_j\Psi_{j+1}+{\rm H.c.}\right]\,.
\end{equation}
In both cases we have defined the pairing as
\begin{equation}
{\bm\Delta}^{\rm s,d}_j=\Delta^{\rm s,d}_j {\rm e}^{\im\kappa_j}\frac{{\bm\tau}_x+\im {\bm\tau}_y}{2}+\Delta^{\rm s,d}_j{\rm e}^{-\im\kappa_j} \frac{{\bm\tau}_x-\im{\bm\tau}_y}{2}\,.
\end{equation}
$t$ is the nearest neighbor hopping, $\mu$ a chemical potential, and $\Delta^{\rm s,d}_j$ the magnitude of the superconducting pairing assumed to be induced by a proximity effect.  The superconducting phase is $\kappa_j$. Additionally there is a magnetic field of strength $B$,
\begin{equation}
H_{\rm B}=B\sum_j\Psi^\dagger_j\vec{n}_j\cdot\vec{{\bm\sigma}}\Psi_j\,,
\end{equation}
which can locally vary its orientation in a direction given by the unit vector $\vec{n}_j$ \cite{Gangadharaiah2011,Kjaergaard2012,Martin2012}. Such a magnetic field could for example be created by considering local magnetic moments of magnetic adatoms on a superconducting surface \cite{Choy2011,Nadj-Perge2013,Klinovaja2013b,Braunecker2013,Vazifeh2013,Kim2014a,Poyhonen2014,Heimes2014}. For  class DIII we must have $B=0$ for time-reversal invariance.

Finally we can have spin-orbit coupling
\begin{equation}
H_{\rm so}=-\sum_j\frac{1}{2}\left[\Psi^\dagger_j(\im\alpha{\bm\sigma}^y+\im\beta{\bm\sigma}^x){\bm\tau}^z\Psi_{j+1}+{\rm H.c.}\right]\,.
\end{equation}
Both Rashba and Dresselhaus terms are present, given in a  low-energy approximation by $\alpha{\bm\sigma}^y$ and $\beta{\bm\sigma}^x$ respectively. In this approximate form there is no difference between the effects of Rashba and Dresselhaus couplings and a global spin rotation around the $z$ direction suffices to align the spin-orbit coupling along, for example, ${\bm\sigma}^y$. Note that in this model all terms except for the magnetic field and spin-orbit coupling are invariant under a global spin rotation. We set $t=\hbar=a=1$, where $a$ is the lattice spacing.

The inhomogeneous magnetic field, which can vary its direction arbitrarily, can be described by a unit vector
\begin{equation}\label{mag}
\vec{n}_j=(\cos\vartheta_j\sin\varphi_j,\sin\vartheta_j\sin\varphi_j,\cos\varphi_j)\,,
\end{equation}
though we will generally focus on examples where
\begin{equation}\label{mag2}
\varphi_j=2\pi k_\varphi(j-1){\rm  and }\vartheta_j=2\pi k_\vartheta(j-1)\,.
\end{equation}
We are interested in two different cases with either spin-orbit coupling or inhomogeneous magnetism, though the composite can also be considered.\cite{Klinovaja2012}

If both wires are assumed to be in the topological regime then the matrix elements of $N_1^\dagger H_cN_2$ which couple the topological edge states to the bulk states are small, and the problem approximately decouples into a bulk and edge part. We are only interested in the Majorana states and so we focus on the effective Hamiltonian
\begin{equation}\label{coupled_ham}
{\bm H}_{\gamma}=\left(\begin{array}{cccc}
-\epsilon_1&0&M_{--}&M_{-+}\\
0&\epsilon_1&M_{+-}&M_{++}\\
M^*_{--}&M^*_{+-}&-\epsilon_2&0\\
M^*_{-+}&M^*_{++}&0&\epsilon_2
\end{array}\right)\,.
\end{equation}
The matrix elements are given by
\begin{equation}
M_{ab}=\langle\xi(a\epsilon_1)|H_c|\xi(b\epsilon_2)\rangle\,,
\end{equation}
where $H_{1,2}\xi(a\epsilon_{1,2})=a\epsilon_{1,2}\xi(a\epsilon_{1,2})$ with $a=\pm1$ and $\lim_{L\to\infty}\epsilon_{1,2}\to0$. We will now explicitly use this last property, setting $\epsilon_{1,2}\approx 0$. The eigenvalues of Eq.~\eqref{coupled_ham} are then
\begin{eqnarray}
\lambda^4-\lambda^2\left(\left|M_{++}\right|^2+\left|M_{+-}\right|^2+\left|M_{-+}\right|^2+\left|M_{--}\right|^2\right)\qquad\\\nonumber
+\left|M_{++}M_{--}-M_{+-}M_{-+}\right|^2=0\,.
\end{eqnarray}
Eq.~\eqref{coupled_ham} can also be written in the Majorana state basis, which appears the same after a substitution of the eigenfunctions. We use both formulations in the following arguments. When all four matrix elements vanish, then all four Majorana states survive. When only the $\Order(\lambda^0)$ term vanishes then only two Majorana states survive, this is a finely tuned example and is not usually seen. The extension to cases where there are more than two zero energy states in the wires is straightforward and we do not give it explicitly.

The overlap can be written for eigenstates $\pm\epsilon_{\nu}$ where  $(\psi^{\nu \pm}_j)^T=\{u^{\nu \pm}_{j\uparrow},u^{\nu \pm}_{j\downarrow},v^{\nu \pm}_{j\downarrow},v^{\nu \pm}_{j\uparrow}\}$ explicitly as
\begin{eqnarray}
M_{ab}&=&-\sum_{i,j;\sigma\sigma'}\frac{[t'_{i,j}]_{\sigma\sigma'}}{2}\big[\left(u^{1a}_{i\sigma}-v^{1a}_{i\sigma}\right)^*\left(u^{2b}_{j\sigma'}+v^{2b}_{j\sigma'}\right)\nonumber\\&&
\quad+\left(u^{1a}_{i\sigma}+v^{1a}_{i\sigma}\right)^*\left(u^{2b}_{j\sigma'}-v^{2b}_{j\sigma'}\right)\big]\\\nonumber
&\equiv&\sum_{i,j}M_{ab}^{ij}\,.
\end{eqnarray}
We have used a coupling of the form
\begin{equation}\label{ladder_ham}
H_c=-\sum_{\nu,i,j}\Psi^\dagger_{\nu i}{\bm t}'_{i,j}{\bm\tau}^z\Psi_{\bar\nu j}\,,
\end{equation}
which obeys particle-hole symmetry and has spin structure ${\bm t}'$.
In order for this matrix element to vanish it is sufficient that there is the same chiral Majorana character on any pair of two coupled sites belonging to the two wires and that ${\bm t}'$ does not break the chiral symmetry. This is one of the most robust ways to make the overlap vanish, though it is not the only possible way. This is clearest for the case where $S_j^y=0$ and $\kappa_j=0$, then the eigenstates of the Hamiltonian are real and the Majoranas are purely real at one end of the wire and imaginary at the other. Therefore all $M^{ij}_{ab}=0$ provided $t'_{i,j}$ is short ranged and does not extend along the entire system. This argument extends directly to all planar spin textures and homogeneous superconducting phases, as the required transformations do not alter $M_{ab}$.

\subsection{Coupling BDI wires}

We can now consider a variety of different scenarios starting with coupling BDI wires.
First, by taking $\vec{n}_j=\{0,0,1\}$ and $\beta_\nu=0$, with $\alpha_1=\pm\alpha_2>0$ we can flip the chiral character of Majoranas in the wire 2 by changing this sign. Note that flipping $\alpha\to-\alpha$ is not the only way of reversing the Majorana character, shifting from $\mu=\to 2-\mu$ has the same effect. By varying the coupling between the wires, $t'_{i,j}$, we find the conditions of destroying the Majorana states. Second, by coupling the two wires at one single point, we can consider two wires of different types joined end to end. By calculating the hybridization of Majorana states in disconnected wires when they are brought together, it is possible to predict under what conditions Majoranas destroy each other or co-exist, and to relate this to the Majorana's character.
The same situation exists if we have $\alpha_\nu=\beta_\nu=0$ and a varying magnetic field confined to a plane, as can be seen in 2d arrays \cite{Wang2014,Sedlmayr2015}.

If $\{H,\V\}\neq0$ then only for fine tuned examples can we have $M_{ab}=0$. Generically the Majoranas will hybridize and gain a finite energy. This follows from the observation that $M^{ij}_{ab}$ is no longer fixed to zero and therefore $M_{ab}=0$ requires either a careful cancellation in the sum over position or a tuning of the coupling $t'_{i,j}$. In this case the full system is in the D class, even though each wire separately is BDI and the coupling can be trivial.

Although it is true to say that Majorana states with the same character will not hybridize, if they have opposite chiral character the overlap {\rm e}mph{can} still be zero, again this requires some fine tuning. For example if $\mu_1=0$ and $\mu_2=2$ with all other quantities being equal between the wires, then the Majorana states of the independent wires will be orientated in opposite ways. Consequently in one wire the state $\V=1$ will be on the left and in the other wire it is the $\V=-1$ state that will be on the left. Provided $t'_{i,j}=t'\delta_{ij}$ we still have $M_{ab}=0$. A small amount of disorder is however enough to couple the states giving $M_{ab}\neq 0$. This is an example of weakly protected topological states in a topologically trivial phase, referred to elsewhere as hidden symmetry \cite{Dumitrescu2015b}. In fact if $\{H,\V\}\neq0$ then $M_{ab}=0$ due to two possible reasons. Firstly there can be an emergent symmetry in the low energy sector in which case effectively we recover $\{\tilde H,\V\}=0$ for an approximate low energy Hamiltonian $\tilde H$, and secondly we can have some form of fine tuning of the coupling or weak topology.

The case where we have both $[H,\T_-]=0$ and $[H,\T_+]=0$ is similar to the simple BDI case where the states now exist in Kramer's pairs protected by the $\T_-$ symmetry. For example a system with d-wave pairing, Rashba spin-orbit coupling, and no magnetic field, i.e.~$H_\nu$ with $B=\Delta^{\rm s}_j=0$ has both $[H,\T_+]=0$ and $[H,\T_+]=0$. It can support multiple Kramer's pairs of Majorana states at each end with a character $\V=\C\T_+={\bm \sigma}^y{\bm \tau}^y$. Breaking $\T_-$ takes us back to the usual BDI case with no Kramer's pairs. Breaking $\T_+$ destroys the symmetry protecting the multiple Majoranas and we can have only a single Kramer's pair at each end of the wire.

We can summarize the findings in the following way. For wires in the BDI class then the Majorana states are protected provided the ladder has the same symmetry and the wires are weakly coupled. Strong coupling can drive the system through a topological phase transition back to a topologically trivial regime.  This is not possible with weak coupling as the topological phase transition requires a closing of the bulk gap. If the symmetry is broken then all the Majorana states are normally gapped, though if the ladder remains topologically nontrivial then a single Majorana at each end survives. 

\subsection{Coupling D and DIII wires}

If we start with wires with D symmetry then, except for very carefully tuned examples, the Majorana states are unprotected and become gapped. The DIII case is essentially similar to the D case.
We remind the reader that the DIII symmetry with mirror symmetry is considered as a subset of BDI (see Sec.~\ref{sec_charac}).

If we have two wires with slowly varying superconducting phases $\kappa_j$, where the variation is confined to the bulk of the wires then we are in the D class. Now as we can locally define a character at a single end of the ladder we should be able to have multiple Majorana states, although we are in D. In fact such a situation has an effective low energy theory in the BDI class and the Majoranas can only survive to the extent that one can neglect the corrections to this effective model which lift it out of the BDI class.

\subsection{Coupling Kitaev chains}

If we consider coupling Kitaev chains rather than full spinfull Hamiltonians then we can consider other possibilities. One can define a time-reversal operator which operates in the sublattice space of the ladder, $\T_-={\bm\lambda}^yK$ where ${\bm\lambda}^{x,y,z}$ are Pauli matrices operating in the ladder subspace. It is then possible to construct DIII ladders from D wires and BDI ladders from BDI wires, additionally we can impose the mirror symmetry in this second case. With the mirror symmetry the many possible Majorana states which can be formed at the end of a wire are protected.  For class D the states are in fact protected by the time-reversal invariance of the DIII ladder, if this is broken then they will become gapped as predicted by the absence of a well defined character.

We now give explicit examples. The full system Hamiltonian can be written as
\begin{equation}
{\bm H}=\left(\begin{array}{cc}
H_1&0\\
0&H_2
\end{array}\right)+{\bm H}_c\,.
\end{equation}
The wires, where we have included longer range hopping $t^\nu_{ij}$ and superconductivity $\Delta^\nu_{ij}{\rm e}^{\im\phi^\nu_{ij}}$, are
\begin{eqnarray}
H_\nu&=&\sum_j\Psi^\dagger_{\nu j}\mu_\nu{\bm\tau}^z\Psi_{\nu j}\\\nonumber&&+\sum_{ij}\Psi^\dagger_{\nu i}\left[
\Delta^\nu_{ij}\left(\begin{array}{cc}
0&{\rm e}^{\im\phi^\nu_{ij}}\\
-{\rm e}^{-\im\phi^\nu_{ij}}&0
\end{array}\right)
-t^\nu_{ij}{\bm\tau}^z\right]\Psi_{\nu j}\,.
\end{eqnarray}
where $\Psi^\dagger_{\nu j}=\{c^\dagger_{\nu j},c_{\nu j}\}$, where $c_{\nu j}^{(\dagger)}$ annihilates (creates) a spinless particle at a site $j$ on wire $\nu$. We also introduce $\Psi^\dagger_{j}=\{\Psi^\dagger_{1 j},\Psi^\dagger_{2 j}\}$. This chain can support two Majorana bound states when its symmetry class is D, and many Majorana bound states when its symmetry class is BDI.

If we consider two BDI wires with the symmetry $\{H_\nu,\V\}=0$ where in this case $\V={\bm\tau}^x$, then provided $\{H,\V\}=0$ all Majorana states are protected and the full system is still BDI. For example
\begin{equation}
H_c=t'\sum_j\Psi^\dagger_j{\bm\lambda}^x{\bm\tau}^z\Psi_j\,.
\end{equation}
A coupling which breaks $\{H,\V\}=0$, or a combination for different BDI wires which breaks this symmetry, ends up in a D ladder with no Majorana states surviving. Note we are by construction in the topologically trivial regime of the D ladder.

We can also construct a coupling which gives us the effective time-reversal invariance, $[H,\T_-]=0$, for $\T_-={\bm\lambda}^yK$, given by
\begin{equation}
H_c=\Delta'\sum_j\Psi^\dagger_j{\bm\lambda}^y{\bm\tau}^z\Psi_j\,.
\end{equation}
If the wires we couple are in the BDI class then this extra symmetry corresponds to mirror symmetry in the ladder Hamiltonian, and we have many Kramer's pairs of Majorana states. If we couple D wires then we have no mirror symmetry and only a Kramer's pair of Majorana states is possible, in agreement with the symmetry class. Note that in this case the Majorana bound states of the uncoupled D chains survive as the Kramer's pair in the DIII ladder, they are protected by the time-reversal invariance. Longer range coupling between the wires does not change these conclusions.

We have extensively checked all the results of the analysis in this section, which is based on the hybridization of the zero energy states inside the gap, by performing an exact scaling analysis on the low energy states as a function of system size. Numerically solving the tight binding Hamiltonian Eq.~(\ref{ladder_ham}) we can unambiguously test whether the Majorana states survive. In all cases, provided inter-wire coupling is weak $t'\lesssim t/5$, the results are all verified.

\section{Braiding characterless Majoranas}\label{braiding}

To understand if the characterless Majoranas ($\{H,\V\}\neq0$ ) would braid differently we consider a generic example of such states and map the corresponding model to the Kitaev model. In the limit of large magnetic fields it is possible to map the spinfull problem to a simpler low-energy model similar to a Kitaev chain \cite{Alicea2011}. First we consider the continuum limit of our model for slowly varying magnetic fields.
After a gauge transformation which fixes the local spin direction to be parallel to the magnetic field, see App.~\ref{app_gauge}, we have a homogeneous Hamiltonian with a vector potential. The vector potential is defined as $\vec{\mathbf{A}}(\vec{r})=\mathbf{\U}^\dagger(\vec{r})\nabla \mathbf{\U}(\vec{r})$. This can be understood as the term which appears in the lowest order expansion of the effective spin-orbit term $\mathbf{S}^{ij}\approx \vec{\delta}\cdot\vec{\mathbf{A}}(\vec{R}_i)$ where $\vec{R}_i$ is the real space position of the 
$i$ site and once again $\vec{\delta}_{ij}$ is the nearest-neighbour hopping vector from $j$ to $i$. Naturally, for the one-dimensional systems considered here, such terms are non-zero only along the $x$-direction. We find
\begin{equation}\label{approx_gauge_pot}
\mathbf{A}_{x}(\vec{R}_j)=\frac{\im}{2}\left[\partial_{x}\varphi{\bm \sigma}^y+\partial_{x}\vartheta\left(\cos\varphi{\bm \sigma}^z-\sin\varphi{\bm \sigma}^x\right)\right]\,.
\end{equation}
The Hamiltonian, Eq.~\eqref{hamiltonian} with $\alpha=\beta=\Delta^{\rm d}_j=\kappa_j=0$ and $\Delta^{\rm s}_j=\Delta$, then becomes
\begin{eqnarray}
H&\approx&\sum_\sigma\int\ud x\psi_{\sigma,x}^\dagger\left[\hat\epsilon-\mu+B\sigma^z_{\sigma\sigma}\right]\psi_{\sigma,x}\nonumber\\&&-\Delta\int\ud x\psi_{\downarrow,x}\psi_{\uparrow,x}+{\rm H.c.}\\\nonumber&&
+\int\ud x\psi_{\uparrow,x}\frac{1}{2}\left[\partial_x\varphi-\im\partial_x\vartheta\sin\varphi\right]\partial_x\psi_{\downarrow,x}+{\rm H.c.}\,.
\end{eqnarray}
$\hat\epsilon{\rm e}quiv-\partial^2_{xx}/(2m)$ and the diagonal contribution to the gauge potential has been neglected. At low energy, and for sufficiently small spin-orbit terms this can be mapped to a spinless model via
\begin{eqnarray}
\psi_{\uparrow,x}&\sim&\frac{\partial_x\varphi-\im\partial_x\vartheta\sin\varphi}{2B}\partial_x\psi_{x}\,{\rm  and}\\\nonumber
\psi_{\downarrow,x}&\sim&\psi_{x}
\end{eqnarray}
resulting in the effective single band model
\begin{eqnarray}
H&\sim&\int\ud x\psi_{x}^\dagger\left[\hat\epsilon-\mu_{\rm eff}\right]\psi_{x}\nonumber\\&&-\int\ud x|\Delta_{\rm eff}|{\rm e}^{-\im\phi_{\rm eff}}\psi_{x}\partial_x\psi_{x}+{\rm H.c.}\,.
\end{eqnarray}
The effective parameters are
\begin{eqnarray}
\mu_{\rm eff}&=&\mu+B\,,\nonumber\\
|\Delta_{\rm eff}|&=&\frac{\Delta}{2B}\left[\partial_x\varphi^2+\partial_x\vartheta^2\sin^2\varphi\right]\,,\,{\rm  and}\\\nonumber
\phi_{\rm eff}&=&\tan^{-1}\frac{\partial_x\vartheta\sin\varphi}{\partial_x\varphi}\,.
\end{eqnarray}
We have an effective Kitaev model with a spatially varying superconducting pairing and phase. 

As an example we consider a combination of a precessing and rotating field, given by Eq.~\eqref{mag2}, which gives rise to characterless Majoranas ($\{H,\V\}\neq0$ ). The effective parameters become
\begin{eqnarray}
|\Delta_{\rm eff}|&=&\frac{2 \pi^2\Delta}{B}\left[k_\varphi^2+k_\vartheta^2\sin^2(2\pi k_\varphi x)\right]\,,\,{\rm  and}\nonumber\\
\phi_{\rm eff}&=&\tan^{-1}\frac{k_\vartheta\sin(2\pi k_\varphi x)}{k_\varphi}\,.
\end{eqnarray}
However such a phase variation plays no role in the braiding properties of the states, and can be deformed continuously without affecting the results, provided the gap is not closed \cite{Ivanov2001,Alicea2011}. Thus we would expect that at a junction of three wires, the braiding operations with such characterless Majorana states remain the same as those predicted for the simpler ($\{H,\V\}=0$) models. Moreover, our observation re-enforces the fact that the braiding of the Majorana states should not be affected by magnetic impurities.

\section{Conclusions}\label{conclusions}

We have investigated the possibility of characterizing the Majorana bound states which occur in topological superconductors.  We find two distinct cases. On the one hand, which we call case (A), we have all systems with a time-reversal symmetry whose anti-unitary operator squares to 1.  This is the BDI class.
We found that this class also  includes time-reversal invariant systems with a mirror symmetry (which was originally considered as a subset of  the DIII class with mirror symmetry \cite{Zhang2013a}).
This includes for example all s-wave and d-wave superconductors with spin confined to a plane and a homogeneous superconducting phase.
In these systems it is possible to find an operator $\V$ which satisfies the anti-commutation relation $\{H,\V\}=0$, and to unambiguously identify a Majorana character, and thus two different types of Majorana states. These are the eigenstates of $\V$ with eigenvalues $\pm1$. States with opposite character are localized and well separated. We have provided several examples of systems in which writing down the $\V$ operator is possible, as well as the explicit form of the Majorana character operator for each example.
It is also possible for the spin degree of freedom to be replaced by a sub-lattice degree of freedom, the same arguments then apply.

For case (B) where no operator $\V$ satisfies $\{H,\V\}=0$, it is impossible to ascribe a character to the Majorana states, these are the D and DIII topological superconductors.
We want to point out however that the labelling of the Majorana states, associated with their character, can be extended to systems where the properties hold at least locally, for example to SNS ring junctions along which the superconducting phase varies slowly, or to SNS junctions with different SC phases in the two SC regions. In as much as the conclusions of the character can still be applied there is a low energy effective theory in a higher symmetry class (BDI).

We have also considered the consequences of the existence of the two types on Majoranas on their hybridization/scattering and braiding properties.  For case (A) it can be shown that states with opposite character can hybridize, destroying each other, whereas states with equivalent character can not hybridize provided $\{H,\V\}=0$ holds. This is in agreement with the expectations from the topological invariant. Case (A) is always in the BDI class. In 1d BDI has a $\mathbb{Z}$ invariant, allowing for multiple Majorana bound states at a single end.
For case (B) hybridization to a finite energy state is the norm, the exceptions being all finely tuned, though not impossible to find, examples. We note that case (B) is in fact the desired behaviour for braiding of Majorana states at junctions of wires. That the existence of multiple Majorana states is not possible is in agreement with the system being described by a $\mathbb{Z}_2$ invariant.

We have also considered the implications of case (B) on the braiding properties of the Majorana states by mapping one characteristic $(\{H,\V\} \ne 0)$ model to a Kitaev chain with inhomogeneous superconducting phase. We find that the non-abelian braiding remains intact. A good confirmation of this would be a numerical analysis of braiding on characterless Majorana bound states.

\acknowledgements

The authors would like to thank Juan-Manuel Aguiar-Hualde for discussions. This work is supported by the ERC Starting Independent Researcher Grant NANOGRAPHENE 256965.

\vspace*{0.5cm}

Just prior to submission we became aware of similar results on the nature of mirror symmetry in Ref.~\onlinecite{Dumitrescu2015} derived by considering the Berry phase and the winding numbers of the topological phases, and to a review of topological classification and symmetries including spatial and mirror symmetries in Ref.~\onlinecite{Chiu2015a}.

\appendix

\section{Planar spin configurations}\label{app_planar}

If we focus on a spinfull Bogoliubov-de Gennes (BdG) Hamiltonian with only real elements, (see for example the Hamiltonian in Eq.~\eqref{hamiltonian} where the superconducting phase is zero and the spin is confined to the $xz$-plane), then $\V$ takes a particularly simple form: $\V={\bm \sigma}^y{\bm \tau}^y$. That $\V$ anti-commutes with the Hamiltonian follows immediately from the particle-hole symmetry and the realness of the Hamiltonian. In this case $\T_+=K$ and therefore  $\V=\C\T_+$ is satisfied.

By rotating from the trivial case we can find the generic operator $\V$. We start with a system with some generic spin density which is confined to a plane.
Consider a global rotation between two orthonormal bases $\tilde\psi=\mathbf{T}\psi$
\begin{equation}\label{trotate}
\mathbf{T}={\rm e}^{-\frac{\im\theta}{2}{\bm \sigma}^z} {\rm e}^{-\frac{\im\phi}{2}{\bm \sigma}^y}{\bm \tau}^0\,.
\end{equation}
We rotate the Majorana polarization from a basis in which the spin is confined to the $xz$-plane,  $\{\tilde\psi_i\}$, to the `original' basis of our problem, $\{\psi_i\}$. Then from $\{\tilde H,\tilde \V\}=0$ with $\tilde \V={\bm \sigma}^y{\bm \tau}^y$ we have $\V=\mathbf{T}^\dagger\tilde \V\mathbf{T}$ resulting in
\begin{equation}\label{fullpx}
\V=\left(\cos[\theta]{\bm\sigma}^y-\sin[\theta]\cos[\phi]{\bm\sigma}^x-\sin[\theta]\cos[\phi]{\bm\sigma}^z\right){\bm\tau}^y
\,,
\end{equation}
acting locally in space (the spatial dependence will be suppressed where it is not explicitly required). The rotations in spin space, along with the change of phase between the particle and hole spaces which follows, define the full 6-d space in which the set of possible $\V$ operators exist. 

An arbitrary but homogeneous superconducting phase of $\kappa$ requires the transformation $\V\to {\bm G}^\dagger \V {\bm G}$ where
\begin{equation}
{\bm G}=\left(\begin{array}{cccc}
{\rm e}^{\im\kappa/2}&0&0&0\\
0&{\rm e}^{\im\kappa/2}&0&0\\
0&0&{\rm e}^{-\im\kappa/2}&0\\
0&0&0&{\rm e}^{-\im\kappa/2}\end{array}\right)
\end{equation}
is the gauge transformation which sets the superconducting phase to zero in the Hamiltonian.

As we have a universal planar spin density then the angles $\theta$ and $\phi$ can be calculated from two, non-parallel, spin vectors at different spatial points. Let these be $\vec S_{1,2}$, then the spin density is confined to the plane perpendicular to the unit vector
 \begin{equation}\label{ns}
 \hat n_s=\frac{\vec S_1\times \vec S_2}{|\vec S_1\times \vec S_2|}\,,
 \end{equation}
 and passing through the point $(0,0,0)$.
 The necessary rotation is that which turns this into $\hat y$, so that the spin density is confined to the $xz$ plane and the Hamiltonian is real. Therefore we can write
  \begin{equation}
 \hat n_s=(\sin\theta\cos\phi,\cos\theta,\sin\theta\sin\phi)\,,
 \end{equation}
 from which we can calculate $\theta$ and $\phi$.
 We can therefore rewrite the operator, Eq.~(\ref{fullpx}), as
  \begin{equation}\label{eqv}
\V=-{\bm \tau}^y\vec{\bm \sigma}^*\cdot\frac{\vec S_1\times \vec S_2}{|\vec S_1\times \vec S_2|}\,,
 \end{equation}
where naturally $\vec{\bm \sigma}^*$ refers to the complex conjugate of $\vec{\bm \sigma}$.
Global changes to the superconducting phase can be trivially incorporated, and we do not add them here explicitly for the sake of clarity.

\section{Gauge transformation for magnetic inhomogeneity}\label{app_gauge}

To understand why a spatially varying magnetic field can generate Majorana states it is helpful to make a gauge transformation and find the spin-orbit like terms which are generated by the
gauge potential.\cite{Haldane1988b}{${}^{,}$}\cite{Braunecker2010}{${}^{,}$}\cite{Nadj-Perge2013}{${}^{,}$}\cite{Klinovaja2013}
This procedure is well known, but we repeat it here for clarity and completeness. This gauge transformation rotates the spin direction to that defined locally by the magnetic field.  In such a basis the particles have an effective spin-orbit like coupling.
A rotation $\Psi_{j\sigma}=\sum_{\sigma'}\U_{j,\sigma\sigma'}\tilde\Psi_{j\sigma'}$ which diagonalizes the magnetic field term such that
\begin{equation}
\mathbf{\U}^\dagger_j\,\vec{n}_j\cdot\vec{{\bm \sigma}}\,\mathbf{\U}_j={\bm \sigma}^z
\end{equation}
is
\begin{equation}\label{gauge}
\mathbf{\U}_j={\rm e}^{-\frac{\im\vartheta_j}{2}{\bm \sigma}^z} {\rm e}^{-\frac{\im\varphi_j}{2}{\bm \sigma}^y}{\bm \tau}^0\,.
\end{equation}
This gauge transformation results in the Hamiltonian
\begin{equation}
H_0+ H_{\rm B}+ H_{\rm so}+ H_{\rm s}\to\tilde H_0+\tilde H_{\rm B}+\tilde H_{\rm so}+\tilde H_{\rm s}\,.
\end{equation}
The last term is generated by the gauge transformation from the hopping in $H_0$. The s-wave pairing term is invariant under such a rotation. We note that d-wave pairing would be modified, a case we do not consider here. By construction $H_{\rm B}\to\tilde H_{\rm B}$, a diagonal homogeneous magnetic field. More explicitly
\begin{eqnarray}
\tilde H_0+\tilde H_{\rm B}+\tilde H_{\rm s}&=&\sum_j\tilde\Psi^\dagger_j\left[(\mu-t){\bm\tau}^z+B{\bm\sigma}^z-\Delta{\bm\tau}^x\right]\tilde\Psi_j\nonumber\\&&
-\frac{t}{2}\sum_{<i,j>}\tilde\Psi^\dagger_i{\bm\tau}^z\cos\varphi_{ij}\cos\vartheta_{ij}\tilde\Psi_{j}\,.\qquad
\end{eqnarray}
The new interesting term is generated in addition to $\tilde H_0$ by the contributions like ${\tilde c}^\dagger_i \U^\dagger_i\U_j{\tilde c}_j$ and is
\begin{equation}
\tilde H_{\rm so}=-\frac{t}{2}\sum_{<i,j>,\sigma\sigma'}{\tilde\Psi}^\dagger_{i\sigma}S^{ij}_{\sigma\sigma'}{\bm\tau}^z{\tilde\Psi}_{j\sigma'}\,,
\end{equation}
where
\begin{eqnarray}\label{eff_so}
\mathbf{S}^{ij}&=&\im{\bm \sigma}^y\sin\varphi^-_{ij}\cos\vartheta^-_{ij}-\im{\bm \sigma}^x\sin\varphi^+_{ij}\sin\vartheta^-_{ij}\nonumber\\
&&+\im{\bm \sigma}^z\cos\varphi^+_{ij}\sin\vartheta^-_{ij}\,,
\end{eqnarray}
and we have defined $\varphi^\pm_{ij}=(\varphi_i\pm\varphi_j)/2$ and $\vartheta^\pm_{ij}=(\vartheta_i\pm\vartheta_j)/2$. In 2-d systems a mapping between simple magnetic  fields varying in the defined manner and spin-orbit coupling is not possible \cite{Sedlmayr2015}. For the 1-d systems we will focus on here it is straightforward.

\end{document}